\documentclass[11pt]{article}
\usepackage{amssymb}
\usepackage{chicagor}
\usepackage{times}

\newtheorem{THEOREM}{Theorem}[section]
\newenvironment{theorem}{\begin{THEOREM} \hspace{-.85em} {\bf :} }%
                        {\end{THEOREM}}
\newtheorem{LEMMA}[THEOREM]{Lemma}
\newenvironment{lemma}{\begin{LEMMA} \hspace{-.85em} {\bf :} }%
                      {\end{LEMMA}}
\newtheorem{COROLLARY}[THEOREM]{Corollary}
\newenvironment{corollary}{\begin{COROLLARY} \hspace{-.85em} {\bf :} }%
                          {\end{COROLLARY}}
\newtheorem{PROPOSITION}[THEOREM]{Proposition}
\newenvironment{proposition}{\begin{PROPOSITION} \hspace{-.85em} {\bf :} }%
                            {\end{PROPOSITION}}
\newtheorem{DEFINITION}[THEOREM]{Definition}
\newenvironment{definition}{\begin{DEFINITION} \hspace{-.85em} {\bf :} \rm}%
                            {\end{DEFINITION}}
\newtheorem{CLAIM}[THEOREM]{Claim}
\newenvironment{claim}{\begin{CLAIM} \hspace{-.85em} {\bf :} \rm}%
                            {\end{CLAIM}}
\newtheorem{EXAMPLE}[THEOREM]{Example}
\newenvironment{example}{\begin{EXAMPLE} \hspace{-.85em} {\bf :} \rm}%
                            {\end{EXAMPLE}}
\newtheorem{REMARK}[THEOREM]{Remark}
\newenvironment{remark}{\begin{REMARK} \hspace{-.85em} {\bf :} \rm}%
                            {\end{REMARK}}

\newcommand{\thm}{\begin{theorem}}
\newcommand{\lem}{\begin{lemma}}
\newcommand{\pro}{\begin{proposition}}
\newcommand{\dfn}{\begin{definition}}
\newcommand{\rem}{\begin{remark}}
\newcommand{\xam}{\begin{example}}
\newcommand{\cor}{\begin{corollary}}
\newcommand{\prf}{\noindent{\bf Proof:} }
\newcommand{\ethm}{\end{theorem}}
\newcommand{\elem}{\end{lemma}}
\newcommand{\epro}{\end{proposition}}
\newcommand{\edfn}{\bbox\end{definition}}
\newcommand{\erem}{\bbox\end{remark}}
\newcommand{\exam}{\bbox\end{example}}
\newcommand{\ecor}{\end{corollary}}
\newcommand{\eprf}{\bbox\vspace{0.1in}}
\newcommand{\beqn}{\begin{equation}}
\newcommand{\eeqn}{\end{equation}}

\newcommand{\bbox}{\vrule height7pt width4pt depth1pt}
\newcommand{\qed}{\eprf}
\newcommand{\clm}{\begin{claim}}
\newcommand{\eclm}{\end{claim}}

\newcommand{\sat}{\models}

\newcommand{\rimp}{\Rightarrow}

\newcommand{\inter}{\cap}

\renewcommand{\phi}{\varphi}

\newcommand{\B}{{\cal B}}

\newcommand{\K}{{\cal K}}
\newcommand{\M}{{\cal M}}

\renewcommand{\P}{{\cal P}}

\newcommand{\W}{{\cal W}}

\newcommand{\Z}{{\cal Z}}

\newcommand{\ol}{\setlength{\itemsep}{0pt}\begin{enumerate}}
\newcommand{\eol}{\end{enumerate}\setlength{\itemsep}{-\parsep}}
\newcommand{\ul}{\setlength{\itemsep}{0pt}\begin{itemize}}
\newcommand{\dl}{\setlength{\itemsep}{0pt}\begin{description}}
\newcommand{\edl}{\end{description}\setlength{\itemsep}{-\parsep}}
\newcommand{\eul}{\end{itemize}\setlength{\itemsep}{-\parsep}}

\newcommand{\true}{{\it true}}

\newcommand{\RAT}{\mathit{RAT}}
\newcommand{\play}{\mathit{play}}

\newcommand{\commentout}[1]{}

\newcommand{\bi}{\begin{itemize}}
\newcommand{\ei}{\end{itemize}}
\newcommand{\be}{\begin{enumerate}}
\newcommand{\ee}{\end{enumerate}}

\newcommand{\SRAT}{SRAT}
\newcommand{\WRAT}{WRAT}
\newcommand{\NSD}{\mathit{NSD}}
\renewcommand{\B}{B}
\renewcommand{\K}{B^*}
\newcommand{\EB}{EB}
\newcommand{\EK}{EB^*}
\newcommand{\CB}{CB}
\newcommand{\CKR}{CCBR}
\newcommand{\CK}{CB^*}
\newcommand{\MKS}{\mathrm{KS}}
\newcommand{\MKW}{\mathrm{KW}}
\newcommand{\MKR}{\mathrm{KR}}

\renewcommand{\M}{M}

\newcommand{\strat}{\mathbf{s}}
\newcommand{\intension}[1]{[\![ #1 ]\!]_M}

\newcommand{\PR}{\mathcal{PR}}
\newcommand{\cPR}{\mathcal{PR}^c}
\renewcommand{\L}{\mathcal{L}}

\newcommand{\Supp}{\mathit{Supp}}

\renewcommand{\Z}{{\cal Z}}

\begin{document}

\title{Game Theory with Translucent Players}

\author{Joseph Y. Halpern and Rafael Pass\thanks{Halpern is supported in part by NSF grants IIS-0812045,
IIS-0911036, and CCF-1214844, by AFOSR grant FA9550-08-1-0266, and by
ARO grant W911NF-09-1-0281. Pass is supported in part by a Alfred P. Sloan Fellowship,
  Microsoft New Faculty Fellowship, NSF Award CNS-1217821, NSF CAREER
 Award CCF-0746990, NSF Award CCF-1214844, AFOSR YIP Award
  FA9550-10-1-0093, and DARPA and AFRL under contract FA8750-11-2-
  0211. The views and conclusions contained in this document are those
  of the authors and should not be interpreted as representing the
  official policies, either expressed or implied, of the Defense
  Advanced Research Projects Agency or the US Government.}\\
Cornell University \\
Department of Computer Science\\
Ithaca, NY, 14853, U.S.A. \\  e-mail:
\{halpern,rafael\}@cs.cornell.edu}
%
%
\maketitle

\begin{abstract}
A traditional assumption in game theory is that players are opaque to
one another---if a player changes strategies, then this change in strategies does not
affect the choice of other players' strategies. In many situations this
is an unrealistic assumption. We develop a framework for
reasoning about games where the players may be \emph{translucent} to one
another; in particular, a player may believe that if she were to change
strategies, then the other player would also change strategies. 
Translucent players may achieve significantly more efficient outcomes
than opaque ones.

Our main result is a characterization of strategies consistent with
appropriate analogues of common belief of rationality.
\emph{Common Counterfactual Belief of Rationality (CCBR)} holds if 
(1) everyone is 
rational, (2) everyone counterfactually believes that everyone
else is rational (i.e., all players $i$ believe
that everyone else would still be rational even if $i$ were to switch
strategies), (3)
everyone counterfactually believes that everyone else is rational, and
counterfactually believes that 
everyone else is rational, and so on.
CCBR characterizes
  the set of strategies surviving iterated removal of
  \emph{minimax dominated} strategies, where a
strategy $\sigma$ for player $i$ is minimax dominated by $\sigma'$ if
the worst-case payoff for $i$ using $\sigma'$ is better than the best
possible payoff using $\sigma$. 
\end{abstract}
\thispagestyle{empty}
%
%

%

\section{Introduction}
Two large firms 1 and 2 need to decide whether to \emph{cooperate (C)} or
\emph{sue (S)} the other firm. Suing the other 
firm always has a
small positive reward, but being sued induces a high penalty
$p$; more precisely, $u(C,C) = (0,0); u(C,S) = (-p, r); u(S,C) = (r,
-p), u(S,S) = (r-p,r-p)$. In other words, we are considering an instance of
the Prisoner's Dilemma.

But there is a catch. Before acting, each firms needs
to discuss their decision with its board. Although these discussions
are held behind closed doors, there is always the possibility of the
decision being ``leaked''; as a consequence, the
other company may change its course of action.
Furthermore, both companies are aware of this fact.
In other words, the players are \emph{translucent} to one another.

In such a scenario, it may well be rational for both companies to
cooperate. For instance, consider the following situation.
\begin{itemize}
\item Firm $i$ believes that its action is leaked
to firm $2-i$ with probability $\epsilon$.
\item Firm $i$ believes that if the
other firm $2-i$ finds out that $i$ is defecting, then $2-i$ will also
defect.
\item Finally, $p\epsilon > r$ (i.e., the penalty for being sued is
significantly
  higher than the reward of suing the other company).
\end{itemize}
Neither firm defects,  since defection is noticed by the other
firm with probability $\epsilon$, which (according to their
beliefs) leads to a harsh 
punishment. Thus, the possibility of the players' actions being
leaked to the other player allows the players to significantly improve 
social welfare in equilibrium. (This suggests that
it may be mutually beneficial for two countries to spy on each other!)

Even if the Prisoner's dilemma is not played by corporations but by
individuals, each player may believe that if he chooses to defect, his
``guilt'' over defecting may be visible to the other
player. (Indeed, facial and bodily cues such as increased
  pupil size are often associated with deception; see e.g., \cite{EkmanFriesen69}.)
Thus, again, the players may choose to cooperate out of fear that if
they defect, the other player may detect it and act on it.

Our goal is to capture this type of reasoning formally.
We take a Bayesian approach:
Each player has a (subjective) probability distribution (describing
the player's beliefs) over the states of the world.
Traditionally, a player $i$ is said to be rational in a state $\omega$ if
the strategy $\sigma_i$ that $i$ plays at $\omega$ is a best response to
the strategy profile $\mu_{-i}$ of the other players induced by $i$'s
beliefs in $\omega$;%
\footnote{Formally, we assume that $i$ has a distribution on states, and
at each state, a pure strategy profile is played;
the distribution on states clearly induces a distribution on strategy
profiles for the players other than $i$, which we denote $\mu_{-i}$.}
that is, 
$u_i(\sigma_i, \mu_{-i}) \geq u_i(\sigma_i', \mu_{-i})$ for
all alternative strategies $\sigma'_i$ for $i$.
In our setting, things are more subtle.
Player $i$ may believe that if she were to switch strategies from
$\sigma_i$ to $\sigma'_i$, then
players other than $i$ might also switch strategies.
We capture this using \emph{counterfactuals}
\cite{Lewis73,Stalnaker68}.%
\footnote{A different, more direct, approach for capturing our original
motivating example would 
be to consider and analyze an extensive-form variant $G'$ of the original
normal-form game $G$ that explicitly models the ``leakage'' of players'
actions in $G$, allows the player to react to these leakage
signals by choosing a new action in $G$, which again may be leaked and
the players may react to, and so on. Doing this is subtle.  We would
need to model how players respond to receiving leaked information, and to
believing that there was a change in plan even if information wasn't
leaked.  To make matters worse, it's not clear 
what it would mean that a player is ``intending'' to perform an action
$a$ if players can revise what they do as the result of a leak.  Does it
mean that a player will do $a$ if no information is leaked to him?  What if
no information is leaked, but he believes that the other side is
planning to change their plans in any case?  In addition, modeling the
game in this way would require a
distribution over leakage signals to be exogenously given (as part
of the description of the game $G'$).  Moreover,  player strategies
would have to be infinite objects, since there is no bound on the
sequence of leaks and responses to leaks.  
In contrast, using counterfactuals, we can directly reason
about the original (finite) game $G$.}
Associated with each state of the world $\omega$, each player $i$, and
$f(\omega,i,\sigma'_i)$ where player $i$ plays $\sigma'_i$.
Note that if $i$ changes strategies, then this change in strategies
may start a chain reaction, leading to further changes.
We can think of $f(\omega,i,\sigma_i')$ as the steady-state outcome of
this process: the state that would result if $i$ switched strategies to
$\sigma_i'$.  
Let $\mu_{f(\omega,i,\sigma_i')}$ be the distribution on
strategy profiles of $-i$ (the players other than $i$) induced by $i$'s
beliefs at 
$\omega$ about the steady-state outcome of this process.
We say that $i$ is rational at a state $\omega$ where $i$ plays
$\sigma_i$ and has beliefs $\mu_i$ if $u_i(\sigma_i,
\mu_{-i}) \geq u_i(\sigma'_i, \mu_{f(\omega,i,\sigma_i')})$ for every
alternative strategy $\sigma'_i$
for $i$.
Note that we have required the closest-state function to be
deterministic, returning a unique state, rather than a distribution over
states.  While this may seem incompatible with the motivating scenario,
it does not seem so implausible in our context that, by taking a
rich enough representation of states, we can assume that a state
contains enough information about players to resolve uncertainty
about what strategies they would use if one player were to switch
strategies.

We are interested in considering analogues to rationalizability in a
setting with translucent players, and providing epistemic
characterizations of them.  To do that, we need some definitions.
We say that a player $i$ \emph{counterfactually believes} $\phi$ at $\omega$
if $i$ believes $\phi$ holds even if $i$ were to switch strategies.
\emph{Common Counterfactual Belief of Rationality (CCBR)} holds if 
(1) everyone is
rational, (2) everyone counterfactually believes that everyone
else  is rational (i.e., all players $i$ believe
that everyone else would still be still rational even if $i$ were to switch
strategies), (3)
everyone counterfactually believes that everyone else is rational, and
counterfactually believes that 
everyone else is rational, and so on.

Our main result is a characterization of strategies consistent with
CCBR. Roughly speaking, these results can be summarized as follows:
\begin{itemize}
\item If the closest-state function respects ``unilateral
  deviations''---when $i$ switches strategies, the strategies and
beliefs of players other than $i$ remain the same---then CCBR characterizes the
  set of rationalizable strategies.
\item If the closest-state function can be arbitrary, CCBR characterizes
  the set of strategies that survive 
iterated removal of \emph{minimax dominated} strategies: a strategy $\sigma_i$ is
  minimax dominated for $i$ if there exists a strategy $\sigma'_i$ for
  $i$ such that $\min_{\mu'_{-i}}u_i(\sigma'_i,\mu'_{-i}) >
  \max_{\mu_{-i}} u_i(\sigma_i, \mu_{-i})$; that is, 
$u_i(\sigma'_i, \mu'_{-i}) > u_i(\sigma_i, \mu_{-i})$ no matter what the
strategy profiles $\mu_{-i}$ and $\mu'_{-i}$ are. 
\end{itemize}
We also consider analogues of Nash equilibrium in our
setting, and show that individually rational strategy profiles that
survive iterated removal of minimax dominated strategies characterize
such equilibria.

Note that in our approach, each player $i$ has a \emph{belief} about how
the other players' strategies would change if $i$
were to change strategies, but we do not require $i$ to explicitly
specify how he would respond to other people changing
strategies. The latter approach, of having each player 
specify how she responds to her opponents' actions, goes back to von
Neumann and Morgenstern \cite[pp. 105--106]{vNM}:
\begin{quote}
\emph{Indeed, the rules of the game $\Gamma$ prescribe that each
  player must make his choice (his personal move) in ignorance of the
  outcome of the choice of his adversary. 
It is nevertheless conceivable that one of the players, say 2, " finds
out"; i.e., has somehow acquired the
knowledge as to what his adversary's strategy is. The basis for this
knowledge does not concern us; it may (but need not) be experience from
previous plays.}
\end{quote}
Von Neumann and Morgenstern's analysis corresponds to a
\emph{single} round of removal of minimax
dominated strategies. This approach was further explored and formalized by 
by Howard \citeyear{Howard} in the
1960s. In Howard's approach, players pick a
``meta-strategy'' that takes as input the strategy of other players.
It led to complex formalisms involving infinite hierarchies
of meta-strategies: at the lowest level, each player specifies a
strategy in the original game; at level $k$, each player specifies
a ``response rule'' (i.e.,  a meta-strategy) to other players'
$(k-1)$-level response rules. 
Such hierarchical structures have not
proven useful when dealing with applications.
Since we do not require players to specify reaction rules, we
avoid the complexities of this approach. 

\emph{Program equilibria} \cite{tennenholtz} and \emph{conditional
commitments} 
\cite{KKLS} provide a different approach to avoiding infinite
hierarchies. Roughly speaking, each player $i$ simply specifies a
\emph{program} $\Pi_i$; player $i$'s action is determined by
running $i$'s program on input the (description of) the programs of
the other players; that is, $i'$ action is given by $\Pi_i(\Pi_{-i})$.
%
%
%
Tennenholtz
\citeyear{tennenholtz} and Kalai et al. \citeyear{KKLS}
show that every (correlated) individually rational outcome can be
sustained in a program equilibrium. Their model, however,
assumes that player's programs (which should be interpreted as their
``plan of action'') are commonly known to all players. We dispense 
with this assumption. It is also not clear how to define common
belief of rationality in their model; the study of 
program equilibria and conditional commitments has considered only
analogues of Nash equilibrium.

Perhaps most closely related to our model is a paper by Spohn
\citeyear{Spohn03} 
that studies a generalization of Nash equilibrium called
\emph{dependency equilibria}, where players' conjectures are described
as ``conditional 
probabilities'': for each action $a_1$ of player 1, player 1 may
have a different belief about the action of player 2. 
Independently of our work, Salcedo \citeyear{Salcedo}, 
defines a notion of \emph{conjectural rationalizability} that replaces
beliefs (over actions) by conjectures described as conditional
probabilities, as in \cite{Spohn03}. Salcedo also defines a notion of
minimax domination 
(which he calls \emph{absolute domination}), and characterizes conjectural
rationalizability in terms of strategies surviving iterated deletion
of minimax dominated strategies.
Counterfactuals have been explored in a game-theoretic setting; see, for
example, 
\cite{Aumann95,Hal19,Samet96,Stalnaker92,Zambrano04}.
However, all these papers considered only structures where,
in the closest state where $i$ changes strategies, all other players'
strategies remain the same; thus, these approaches are not applicable in our
context.
%
%
%

%
%

%
%
%
\section{Counterfactual Structures}\label{sec:Kripke}
Given a game $\Gamma$, let $\Sigma_i(\Gamma)$ denote player $i$'s pure
strategies in $\Gamma$ (we occasionally omit the parenthetical $\Gamma$
if it is clear from context or irrelevant). 

%

To reason about the game $\Gamma$, we consider a class of
Kripke structures corresponding to $\Gamma$.
For simplicity, we here focus on finite structures. %
A \emph{finite probability structure $M$ appropriate for $\Gamma$} 
is a tuple $(\Omega, \strat,\PR_1, \ldots, \PR_n)$, where
$\Omega$ is a finite set of  
states; $\strat$ associates with each state $\omega \in \Omega$ a
pure strategy profile $\strat(\omega)$ in the game $\Gamma$; 
and, for each
player $i$, $\PR_i$ 
is a \emph{probability assignment} that
associates with each state $\omega \in \Omega$ a
probability 
distribution $\PR_i(\omega)$ on $\Omega$, such that
\begin{enumerate}
\item $\PR_i(\omega)(\intension{\strat_i(\omega)}) 
= 1$, where
for each strategy $\sigma_i$ for player $i$,
$\intension{\sigma_i} = \{\omega: \strat_i(\omega) = \sigma_i\}$,
where $\strat_i(\omega)$ denotes player $i$'s strategy in the strategy 
profile $\strat(\omega)$;
\item $\PR_i(\omega)(\intension{\PR_i(\omega),i}) = 1$,
where for each probability measure $\pi$ and player $i$, $\intension{\pi,i}
=\{\omega : \PR_i(\omega) = \pi\}$.
\end{enumerate}
These assumptions say that player $i$ assigns probability 1 to his
actual strategy and beliefs.

To deal with counterfactuals, we augment probability
structures with a ``closest-state'' function $f$ that associates with
each state $\omega$, player $i$, and strategy $\sigma'_i$, a state
$f(\omega,i,\sigma_i)$ where
player $i$ plays $\sigma'$; if $\sigma'$ is already played in $\omega$, then the closest state to
$\omega$ where $\sigma'$ is played is $\omega$ itself.  Formally,
a \emph{finite counterfactual structure $M$ appropriate for $\Gamma$} 
is a tuple $(\Omega, \strat,f, \PR_1, \ldots, \PR_n)$, where
$(\Omega, \strat,\PR_1, \ldots, \PR_n)$ is a probability structure
appropriate for $\Gamma$ and 
$f$ is a ``closest-state'' function.  We require that if 
$f(\omega,i,\sigma'_i) = \omega'$, then
\begin{enumerate}
\item $\strat_i(\omega') =
  \sigma'$; 
\item if $\sigma'_i = \strat_i(\omega)$, then $\omega' = \omega$.
\end{enumerate}

%
%
%
%
%
%
%
Given a probability assignment $\PR_i$ for player $i$, we define $i$'s
counterfactual belief at state $\omega$ (``what $i$ believes would happen
if he 
switched to $\sigma'_i$ at $\omega$) as 
$$\cPR_{i,\sigma'_i}(\omega) (\omega') = 
\sum_{\{\omega''\in \Omega :  f(\omega'',i,\sigma'_i) = \omega'\}}
\PR_i(\omega)(\omega'').$$ 
Note that the conditions above imply that each player $i$
knows what strategy he would play if he were to switch; that
is, $\cPR_{i,\sigma'_i}(\omega)(\intension{\sigma'_i}) = 1$.

Let $\Supp(\pi)$ denote the support of the probability measure $\pi$.
Note that $\Supp(\cPR_{i,\sigma'_i}(\omega)) =
\{f(\omega',i,\sigma_i'): \omega' \in \Supp(\PR_i(\omega)\}$. 
Moreover, it is almost immediate from the definition that  if 
$\PR_i(\omega) = \PR_i(\omega')$, then 
$\cPR_{i,\sigma'_i}(\omega) = \cPR_{i,\sigma'_i}(\omega')$ for all
strategies $\sigma'_i$ for player $i$.  
But it does \emph{not} in
general follow that 
$i$ knows his counterfactual beliefs at $\omega$, that is, 
it may not be the case that for all strategies $\sigma'_i$ for
player $i$,
$\cPR_{i,\sigma'_i}(\omega)(\intension{\cPR_{i,\sigma'_i}(\omega),i}) = 1$.  
Suppose that we think of a state as representing each
player's \emph{ex ante} view of the game.  The fact that player
$\strat_i(\omega) = \sigma_i$ should then be interpreted as ``$i$
\emph{intends} to play $\sigma_i$ at state $\omega$.''  With this view,
suppose that $\omega$ is a state where $\strat_i(\omega)$ is a
conservative strategy, while $\sigma_i'$ is a rather reckless strategy.
It seems reasonable to expect that $i$'s subjective beliefs regarding
the likelihood of various outcomes 
may depend in part on whether he is in a conservative or reckless frame
of mind.    We can think of
$\cPR_{i,\sigma'_i}(\omega)(\omega')$ as the probability that $i$
ascribes, at state $\omega$, to $\omega'$ being the outcome of $i$
switching to strategy $\sigma_i'$; thus,
$\cPR_{i,\sigma'_i}(\omega)(\omega')$ represents $i$'s evaluation of the
likelihood of $\omega'$ when he is in a conservative frame of mind.
This may not be the evaluation that $i$ uses in states in the support 
$\cPR_{i,\sigma'_i}(\omega)$; at all these states, $i$ is in a
``reckless'' frame of mind.  Moreover, there may not be a unique reckless
frame of mind, so that $i$ may not have the same beliefs at all the
states in the support of $\cPR_{i,\sigma'_i}(\omega)$.

$M$ is a \emph{strongly appropriate counterfactual structure} if it
is an appropriate counterfactual structure and, at every state
$\omega$, every player $i$
knows his counterfactual beliefs. 
As the example above suggests,
strong 
appropriateness is a nontrivial 
requirement. 
As we shall see, however, our characterization results hold in both
appropriate and strongly appropriate
counterfactual structures.

Note that even in strongly appropriate counterfactually structures, we
may not have 
$\PR_i(f(\omega,i,\sigma_i')) =
\cPR_{i,\sigma'_i}(\omega)$.    
We do have $\PR_i(f(\omega,i,\sigma_i')) =
\cPR_{i,\sigma'_i}(\omega)$ in strongly appropriate counterfactual
structures if $f(\omega,i,\sigma_i')$ is in the support of
$\cPR_{i,\sigma'_i}(\omega)$ (which will certainly be the case if
$\omega$ is in the support of $\PR_i(\omega)$). 
To see why we may not want to have 
$\PR_i(f(\omega,i,\sigma_i')) =
\cPR_{i,\sigma'_i}(\omega)$ in general, even in strongly appropriate
counterfactual structures, consider the example above again.  Suppose that, in
state $\omega$, 
although $i$ does not realize it, he has been given a drug that affects
how he evaluates the state.  He thus ascribes probability 0 to
$\omega$.  In $f(\omega,i,\sigma_i')$ he has also been given the drug,
and the drug in particular affects how he evaluates outcomes.  Thus, $i$'s
beliefs in the state $f(\omega,i,\sigma_i')$ are quite different from
his beliefs in all states in the support of $\cPR_{i,\sigma_i'}(\omega)$.

%
%
%

%
%
%
%

%
\subsection{Logics for Counterfactual Games}
Let $\L(\Gamma)$ be the language where we start with $\true$ and the 
primitive proposition $\RAT_i$ and
$\play_i(\sigma_i)$ for $\sigma_i \in \Sigma_i(\Gamma)$,
and close off under the modal operators
$\B_i$ (player $i$ believes) and $\K_i$ 
(player $i$ counterfactually believes) for 
$i = 1, \ldots, n$, 
$\CB$ (common belief), and $\CK$ (common counterfactual belief),
conjunction, and negation.   
We think of $\B_i \phi$ as saying that ``$i$ believes $\phi$ holds with probability 1''
and $\K_i \phi$ as saying ``$i$ believes that $\phi$
holds with 
probability 1, even if $i$ were to switch strategies''. 

Let $\L^0$ be defined exactly like $\L$ except that we exclude the
``counterfactual'' modal operators $\B^*$ and $\CB^*$.
We first define semantics for $\L^0$ using probability structures
(without counterfactuals).
We define the notion of a formula $\phi$ being true at a state $\omega$ in 
a probability structure $M$ (written $(M,w) \sat \phi$) in the standard
way, by induction on the structure of $\phi$, as follows:
\begin{itemize}
\item $(M,\omega) \sat \true$ (so $\true$ is vacuously true).
\item $(M,\omega) \sat \play_i(\sigma_i)$ iff $\sigma_i = \strat_i(\omega)$.
\item $(M,\omega) \sat \neg \phi$ iff $(M,\omega) \not\sat
\phi$.  
\item $(M,\omega) \sat \phi \land \phi'$ iff $(M,\omega) \sat \phi$ and 
$(M,\omega) \sat \phi'$.
\item $(M,\omega) \sat \B_i \phi$ iff $\PR_i(\omega)(\intension{\phi}) = 1$, where
$\intension{\phi} = \{\omega: (M,\omega) \sat \phi\}$.
\item $(M,\omega) \sat \RAT_i$ iff $\strat_i(\omega)$ is a best
  response given player $i$'s beliefs regarding the strategies of other players induced by
$\PR_i$.
\item Let $\EB \phi$ (``everyone believes $\phi$'') be an abbreviation of
$\B_1 \phi \land \ldots \land \B_n 
\phi$; and define $\EB^{k}
\phi$ for all $k$ inductively, by taking $\EB^1 \phi$ to be $\EB \phi$ and
$\EB^{k+1} \phi$ to be $\EB(\EB^k \phi)$. 
\item $(M,\omega) \sat \CB \phi$ iff $(M,\omega) \sat \EB^k \phi$ for
all $k \ge 1$.
\end{itemize}
Semantics for $\L^0$ in counterfactual structures is defined in an
identical way, except that we redefine $\RAT_i$ to take into account
the fact that player $i$'s beliefs about the strategies of players
$-i$ may change if $i$ changes strategies.
\begin{itemize}
\item $(M,\omega) \sat \RAT_i$ iff for every strategy $\sigma'_i$ for
  player $i$, 
  \begin{eqnarray*}
\sum_{\omega' \in \Omega} \PR_i(\omega)(\omega') u_i( \strat_i(\omega),
\strat_{-i}(\omega'))\geq 
\sum_{\omega' \in \Omega} \cPR_{i,\sigma'_i}(\omega)(\omega') u_i( \sigma'_i,
\strat_{-i}(\omega')).
\end{eqnarray*}
\end{itemize}
The condition above is equivalent to
requiring that
  \begin{eqnarray*}
\sum_{\omega' \in \Omega} \PR_i(\omega)(\omega') u_i( \strat_i(\omega),
\strat_{-i}(\omega'))\geq 
\sum_{\omega' \in \Omega} \PR_i(\omega)(\omega') u_i( \sigma'_i,
\strat_{-i}(f(\omega', i, \sigma'_i))).
\end{eqnarray*}
Note that, in general, this condition is different from requiring that 
$\strat_i(\omega)$ is a best response given player $i$'s
beliefs regarding the strategies of other players induced by
$\PR_i$.

To give the semantics for $\L$ in counterfactual structures, we now also
need to define the semantics of $\K_i$ and $\CK$:
\begin{itemize}
\item $(M,\omega) \sat \K_i \phi$ iff for all strategies $\sigma'_i \in
  \Sigma_{i}(\Gamma)$, 
$\cPR_{i,\sigma'_i}(\omega)(\intension{\phi}) = 1$.
\item $(M,\omega) \sat \CK \phi$ iff $(M,\omega) \sat (\EK)^k \phi$ for
all $k \ge 1$.
\end{itemize}
It is easy to see that, like $B_i$, $B_i^*$ depends only on $i$'s
beliefs; as we  observed above, if $\PR_i(\omega) = \PR_i(\omega')$,
then $\cPR_{i,\sigma'_i}(\omega) = \cPR_{i,\sigma'_i}(\omega')$ for all
$\sigma'_i$, so $(M,\omega) \sat \K_i\phi$ iff $(M,\omega') \sat
\K_i\phi$.  It immediately follows that $\K_i \phi \rimp  \B_i \K_i
\phi$ is valid (i.e., true at all states in all structures).

The following abbreviations will be useful in the sequel.
Let $\RAT$ be an abbreviation for $\RAT_1 \land \ldots \land \RAT_n$,
and
let $\play(\vec{\sigma})$ be an abbreviation for $\play_1(\sigma_1) \land
\ldots \land \play_n(\sigma_n)$.

\subsection{Common Counterfactual Belief of Rationality}
We are interested in analyzing strategies being played at states
where (1) everyone is
rational, (2) everyone counterfactually believes that everyone
else  is rational (i.e., for every player $i$, $i$ believes
that everyone else would still be rational even if $i$ were to switch
strategies), (3)
everyone counterfactually believes that everyone else is rational, and
counterfactually believes that 
everyone else is rational, and so on.
For each player $i$, define the formulas $\SRAT_i^k$ (player $i$ is
strongly $k$-level rational) inductively, by taking
$\SRAT^0_i$ to be $\true$ and $\SRAT^{k+1}_i$ 
to be an abbreviation of 
$$\RAT_i \land \K_i (\land_{j \neq i}\SRAT^k_{j}).$$
Let $\SRAT^k$ be an abbreviation of $\land_{j=1}^n \SRAT^k_{j}$.

Define $\CKR$ (common counterfactual belief of rationality) as
follows: 
\begin{itemize}
\item
$(M,\omega) \sat \CKR$ iff $(M,\omega) \sat \SRAT^k \phi$ for
all $k \ge 1$.
\end{itemize}
Note that it is critical in the definition of $\SRAT^k_i$ that we require
only that player $i$ counterfactually believes that everyone
else (i.e., the players other than $i$) are rational, and believe
that everyone else is rational, and so on. Player
$i$ has no reason to believe that his own strategy would be rational if
he were to switch strategies; indeed, $\K_i \RAT_i$ can hold only if
\emph{every} 
strategy for player $i$ is rational with respect to $i$'s beliefs.
This is why we do not define $\CKR$ as $\CK \RAT$.%
\footnote{Interestingly, Samet \citeyear{Samet96} essentially
considers an analogue  of $\CK \RAT$.  This works in his setting since he
is considering only events in the past, not events in the future.}

We also consider the consequence of just common belief of
rationality in our setting. Define $\WRAT_i^k$ (player $i$ is 
weakly $k$-level rational) just as $\SRAT_i^k$,
except that $\K_i$ is replaced by $\B_i$.
An easy induction on $k$ shows that $\WRAT^{k+1}$ implies
$\WRAT^{k}$ and that $\WRAT^k$  implies $\B_i (\WRAT^k)$.%
\footnote{We can also show that $\SRAT^{k+1}$ implies
$\SRAT^{k}$, but it is not the case that $\SRAT_i^k$ implies $\K_i
\SRAT_i^k$, since $\RAT$ does not imply $\K_i \RAT$.} It follows that we
could have equivalently defined $\WRAT_i^{k+1}$ as 
$$\RAT_i \land B_i (\land_{j=1}^n\WRAT^{k}_{j}).$$ Thus,
$\WRAT^{k+1}$ is equivalent to $\RAT \land \EB (\WRAT^{k})$. 
As a consequence we have the following:
\pro \label{claim1} $(M,\omega) \sat \CB (\RAT)$ iff $(M,\omega) \sat
\WRAT^k$ for all $k \geq 0$.
\epro

\section{Characterizing Common Counterfactual Belief of
  Rationality}\label{sec:strongrat} 

To put our result into context, we first
restate the characterizations of rationalizability given by Tan and
Werlang \citeyear{TW88} and Brandenburger and Dekel \citeyear{BD87a} 
in our language. 
We first recall Pearce's \citeyear{Pearce84} definition of rationalizability.

\dfn \label{rat1} 
A strategy $\sigma_i$ for player $i$ is \emph{rationalizable} if, for
each player $j$, there is a set $\Z_j   
\subseteq \Sigma_j(\Gamma)$ and, for each strategy $\sigma'_j \in \Z_j$,  a   
probability measure $\mu_{\sigma'_j}$ on $\Sigma_{-j}(\Gamma)$ whose
support is a subset of
$\Z_{-j}$ such that     
\begin{itemize}   
\item $\sigma_i \in \Z_i$; and
\item for strategy $\sigma'_j \in \Z_j$, 
strategy $\sigma'_j$ is a best response to (the beliefs)
$\mu_{\sigma'_j}$.
\end{itemize}
A strategy profile $\vec{\sigma}$ is rationalizable if every strategy
$\sigma_i$ in the profile is rationalizable.
\edfn

\thm\label{thm:CBRclassic} {\rm \cite{BD87a,TW88}}
$\vec{\sigma}$ is 
rationalizable 
in a game $\Gamma$ iff there exists a 
finite probability structure  $M$ that is appropriate for
$\Gamma$ and a state $\omega$ such 
that 
$(M,\omega) \sat \play(\vec{\sigma}) \land \CB (\RAT)$.
\ethm

We now consider counterfactual structures.
We here provide a condition on the closest-state function
under which common (counterfactual) belief of rationality 
characterizes rationalizable strategies.

\subsection{Counterfactual Structures Respecting Unilateral Deviations}
Let $M = (\Omega, f, \PR_1, \ldots, \PR_n)$ be a finite counterfactual
structure that 
is 
appropriate for $\Gamma$. $M$ {\em
respects unilateral 
deviations} if, for every state $\omega \in \Omega$, player $i$, and
strategy $\sigma'_i$ for player 
$i$, $\strat_{-i}(f(\omega,i,\sigma')) = \strat_{-i}(\omega)$ and
$\PR_{-i}(f(\omega,i,\sigma')) = \PR_{-i}(\omega)$; that is, in the
closest state to $\omega$ where player 
$i$ switches strategies, everybody else's strategy and beliefs
remain same.

Recall that $\L^0$ is defined exactly like $\L$ except that we exclude the
``counterfactual'' modal operators $\B^*$ and $\CB^*$.
The following theorem shows that for formulas in $\L^0$, counterfactual
structures respecting unilateral deviations behave just as 
(standard) probability
structures. 
\thm\label{thm:L0} 
For every $\phi \in \L^0$, there exists a 
finite probability structure $M$ appropriate for $\Gamma$ and a
state $\omega$ such that $(M,\omega) \sat \phi$ iff there exists a
finite counterfactual structure $M'$ (strongly) appropriate for $\Gamma$
that respects 
unilateral deviations, and a state $\omega'$ such that $(M',\omega')
\sat \phi$.
\ethm
\prf
For the ``if'' direction, let $M' = (\Omega, f, \PR_1, \ldots,
\PR_n)$ be a finite counterfactual structure that is counterfactually appropriate
for $\Gamma$ (but not necessarily strongly counterfactually
appropriate) and respects
unilateral deviations.
Define 
$M= 
(\Omega, \PR_1, \ldots,
\PR_n)$. Clearly $M$ is a finite probability structure appropriate for
$\Gamma$; it 
follows by a 
straightforward induction on the length of $\phi$ that $(M',\omega)
\sat \phi$ iff $(M,\omega) \sat \phi$.

For the ``only-if'' direction, let 
$M = (\Omega, \PR_1, \ldots,
\PR_n)$ be a finite 
probability structure, and let $\omega \in \Omega$ be a state
such that $(M,\omega) \sat \phi$.
We assume without loss of generality that for each strategy profile
$\vec{\sigma}'$ there exists some state $\omega_{\vec{\sigma}'} \in
\Omega$ such that $\strat(\omega_{\vec{\sigma}'}) = \vec{\sigma}'$ and 
for each player $i$,
$\PR_i(\omega_{\vec{\sigma}'})(\omega_{\vec{\sigma}'}) = 1$.
(If such a state does not exist, we can always add it.)

We define a finite counterfactual structure $M' = (\Omega', f', \PR'_1, \ldots,
\PR'_n)$ as follows:
\begin{itemize}
\item $\Omega' = \{ (\vec{\sigma}',\omega'): \vec{\sigma}'
  \in \Sigma(\Gamma), \omega' \in \Omega\}$;
\item $\strat'( \vec{\sigma}', \omega') = \vec{\sigma}'$;
\item $f( (\vec{\sigma}',\omega'), i, \sigma''_i) = ((\sigma''_i,
  \sigma'_{-i}), \omega')$
\item $\PR'_i$ is defined as follows.
 \begin{itemize}
\item $\PR'_i (\strat(\omega'), \omega')( \strat(\omega''), \omega'') = \PR_i(\omega')(\omega'')$
\item If $\vec{\sigma}' \neq \strat(\omega')$,
$\PR'_i (\vec{\sigma}', \omega') (\vec{\sigma}',\omega_{\vec{\sigma}'})
  = 1.$
\end{itemize}
\end{itemize}
It follows by construction that $M'$ is strongly 
appropriate
for $\Gamma$ and respects unilateral deviations.
Furthermore, it follows by an easy induction on the length of the
formula $\phi'$ that for every state $\omega \in \Omega$, $(M,\omega)
\sat \phi'$ iff $(M', (\strat(\omega), \omega)) \sat \phi'$.
\eprf

We can now use Theorem \ref{thm:L0} together with the standard
characterization of common belief of rationality (Theorem
\ref{thm:CBRclassic}) to characterize both common belief of
rationality and common counterfactual belief of rationality.
\thm\label{thm:CBRclassic2} 
The following are equivalent: 
\begin{enumerate}
\item[(a)] $\vec{\sigma}$ is rationalizable in
  $\Gamma$;
\item[(b)] there exists a finite counterfactual structure $M$
that is 
appropriate for $\Gamma$ and respects unilateral deviations, and a state $\omega$ such that
$(M,\omega) \sat \play(\vec{\sigma}) \land_{i=1}^n \WRAT_i^k$ for
all $k \geq 0$; 

\item[(c)] there exists a finite counterfactual structure $M$
that is strongly 
appropriate for $\Gamma$ and respects
unilateral deviations and a state $\omega$ such that 
$(M,\omega) \sat \play(\vec{\sigma}) \land_{i=1}^n \WRAT_i^k$ for
all $k \geq 0$;

\item[(d)] there exists a finite counterfactual structure $M$
that is 
appropriate for $\Gamma$ and respects
unilateral deviations and a state $\omega$ such that 
$(M,\omega) \sat \play(\vec{\sigma}) \land_{i=1}^n \SRAT_i^k$ for
all $k \geq 0$; 

\item[(e)] there exists a finite counterfactual structure $M$
that is strongly 
appropriate for $\Gamma$ and respects
unilateral deviations and a state $\omega$ such that 
$(M,\omega) \sat \play(\vec{\sigma}) \land_{i =1}^n \SRAT_i^k$ for
all $k \geq 0$.
\end{enumerate}
\ethm

\prf 
The equivalence of (a), (b), and (c) is immediate from Theorem
\ref{thm:CBRclassic}, Theorem \ref{thm:L0}, and Proposition~\ref{claim1}.
We now  prove the equivalence of (b) and (d).  
Consider an counterfactual structure $M$ that is 
appropriate for 
$\Gamma$ and respects unilateral deviations.
The result follows immediately once we show
that for all states $\omega$ and all $i \ge 0$, $(M,\omega)
\sat \WRAT^k_i$ iff $(M,\omega) \sat 
\SRAT^k_i$.  An easy induction on $k$ shows that $\SRAT^k_i \rimp
\WRAT^k_i$ is valid in all counterfactual structures, not just ones
that respect unilateral deviations.  We prove the converse in structures
that respect unilateral deviations by induction on $k$.
The base case holds trivially.  For the induction 
step, suppose that $(M,\omega) \sat
\WRAT^k_i$; that is, $(M,\omega) \sat \RAT_i \land \B_i (\land_{j \neq
  i} \WRAT^{k-1}_{j})$.
Thus, for all $\omega' \in \Supp(\PR_i(\omega))$, we have that
$(M,\omega') \sat \land_{j \neq
i} \WRAT^{k-1}_{j}$. Thus, by the induction hypothesis, $(M,\omega')
\sat 
\land_{j \neq 
i} \SRAT^{k-1}_{j}$. Since, as we have observed, the truth of a formula of
the form $\K_j\phi$ at a state $\omega''$ depends only on $j$'s beliefs at
$\omega''$ and the truth of $\RAT_j$ depends only on $j$'s strategy and
beliefs at $\omega''$, it easily follows that, if $j$ has the same
beliefs and plays the same strategy at $\omega_1$ and $\omega_2$, then 
$(M,\omega_1) \sat \SRAT^{k-1}_j$ iff $(M,\omega_2) \sat \SRAT^{k-1}_j$.   
Since $(M,\omega') \sat \land_{j \neq i} \SRAT^{k-1}_{j}$ and 
$M$ respect unilateral deviations, for all strategies $\sigma'_i$,
it follows that $(M,f(\omega', i, \sigma_i')) \sat
\land_{j \neq
  i} \SRAT^{k-1}_{j}$. 
Thus, $(M,\omega) \sat \RAT_i \land \K_i(
\land_{j \neq   i} \SRAT^{k-1}_{j})$,
as desired.  The argument that (c) is equivalent to (e) is identical; we
just need to consider strongly 
appropriate counterfactual structures
rather than just 
appropriate counterfactual structures.
\eprf

\begin{remark}
\label{unilateral.rem}
Note that, in the proofs of Theorems \ref{thm:L0} and
\ref{thm:CBRclassic2}, a weaker condition on the counterfactual
structure would suffice, namely, that we restrict to counterfactual structures 
where, for every state $\omega \in \Omega$, player $i$, and
strategy $\sigma'_i$ for player 
$i$, the projection of $\cPR_{i,\sigma'_i}(\omega)$ onto strategies
and beliefs of players $-i$ is equal to the projection of $\PR_i(\omega)$
onto strategies and beliefs of players $-i$.
That is, every player's counterfactual beliefs 
regarding other players' strategies and beliefs 
are the same as the player's actual beliefs.
\end{remark}

%
%
%
%
%
%
%
%
%
%

%
%
\subsection{Iterated Minimax Domination}
We now characterize common counterfactual belief of
rationality without 
putting any restrictions on the counterfactual structures (other than
them being appropriate, or strongly appropriate).
Our characterization is based on ideas that come from the
characterization of rationalizability.
It is well known that rationalizability can be characterized in
terms of an iterated deletion procedure, where at each stage, a
strategy $\sigma$ for player $i$ is deleted if there are no beliefs that
$i$ could have about the undeleted strategies for the players other than
$i$ that would make $\sigma$ rational \cite{Pearce84}.  Thus, there is a
deletion 
procedure that, when applied repeatedly, results in only the
rationalizable strategies, that is, the strategies that are played in states
where there is common belief of rationality, being left undeleted.
We now show that there is an analogous way of characterizing common
counterfactual belief of rationality.

The key to our characterization is the notion of
\emph{minimax dominated} 
strategies.  

\dfn Strategy $\sigma_i$ for player $i$ in game $\Gamma$ is
\emph{minimax dominated  
with respect to $\Sigma' _{-i}\subseteq \Sigma_{-i}(\Gamma)$} iff there
exists a strategy $\sigma'_i \in \Sigma_i(\Gamma)$ such that  
$$\min_{\tau_{-i} \in
\Sigma'_{-i}} u_i(\sigma'_i, \tau_{-i}) > \max_{\tau_{-i} \in
\Sigma'_{-i}} u_i(\sigma_i, \tau_{-i}).$$
\edfn

In other words, player $i$'s strategy $\sigma$ is minimax dominated
with respect to $\Sigma'_{-i}$ iff there exists
a strategy $\sigma'$ such that the worst-case payoff for player $i$ if
he uses $\sigma'$ is strictly better than his best-case payoff if he
uses $\sigma$, given that the other players are restricted to using a
strategy in $\Sigma'_{-i}$.

In the standard setting, if a strategy $\sigma_i$ for player $i$ is
dominated by $\sigma_i'$ then we would expect that a rational player
will never player $\sigma_i$, because $\sigma_i'$ is a strictly better
choice.  As is well known, if $\sigma_i$ is dominated by $\sigma_i'$,
then there are no beliefs that $i$ could have regarding the strategies
used by the other players according to which $\sigma_i$ is a best response
\cite{Pearce84}.  
This is no longer the case in our setting.  For example,
in the standard setting, cooperation is dominated by defection in
Prisoner's Dilemma. But in our setting, suppose that 
player 1 believes that if he cooperates, then the other player will
cooperate, while if he defects, then the other player will defect.
Then cooperation is not dominated by defection.  

So when can we guarantee that playing a strategy is irrational in our
setting?  This is the case only if the strategy is minimax dominated.
If $\sigma_i$ is minimax dominated by $\sigma_i'$, there are no
counterfactual beliefs that $i$ could have that would justify playing
$\sigma_i$.  Conversely, if $\sigma_i$ is not minimax dominated by any
strategy, then there are beliefs and counterfactual beliefs that $i$
could have that would justify playing $\sigma_i$.  Specifically, $i$
could believe that the players in $-i$ are playing the strategy profile
that gives $i$ the best possible utility when he plays $\sigma_i$, and
that if he switches to another strategy $\sigma_i'$, the other players
will play the strategy profile that gives $i$ the worst possible utility
given that he is playing $\sigma_i'$.  
Note that we consider only domination by pure strategies.   It is easy to
construct examples of strategies that are not minimax dominated by any pure
strategy, but are minimax dominated by a mixed strategy.  Our
characterization works only if we restrict to domination by pure strategies.
The characterization, just as with the characterization of
rationalizability, involves iterated deletion, but now we do not delete
dominated strategies in the standard sense, but minimax dominated strategies.

\dfn Define $\NSD_j^k(\Gamma)$ inductively: let $\NSD_j^0(\Gamma) =
\Sigma_j$ and let $\NSD_j^{k+1}(\Gamma)$
consist of the strategies in $\NSD_j^{k}(\Gamma)$ not minimax 
dominated with respect to $\NSD_{-j}^{k}(\Gamma)$.
Strategy $\sigma$ \emph{survives $k$ rounds of 
iterated deletion of minimax dominated strategies for player $i$} if $\sigma \in
\NSD_i^k(\Gamma)$.
Strategy $\sigma$ for player $i$ survives iterated deletion
of minimax 
dominated strategies if it survives $k$ rounds
of iterated deletion of strongly dominated for all $k$, that is, if 
$\sigma \in \NSD^\infty_i(\Gamma) = \inter_k \NSD^k_i(\Gamma)$.
\edfn

In the deletion procedure above, at each step we remove \emph{all}
strategies that are minimax dominated; that is we perform a ``maximal''
deletion at each step. As we now show, the set of
strategies that survives iterated deletion is actually independent of
the deletion order.
Let $S^0, \ldots, S^m$ be sets of strategy profiles.  
$\vec{S}= (S^0, S^1, \ldots, S^m)$ is a \emph{terminating deletion
sequence} for $\Gamma$ if, for $j = 0, \ldots, m-1$, $S^{j+1} \subset
S^j$ 
(note that we use $\subset$ to mean proper subset)
and all players $i$, $S^{j+1}_i$ contains all strategies
for player $i$ not minimax dominated with respect to $S^j_{-i}$ (but may
also contain some strategies that are minimax dominated),
and $S^m_i$ does not contain any strategies that are minimax dominated
with respect to $S^{m}_{-i}$.
A set $T$ of strategy profiles has \emph{ambiguous}
terminating sets if there exist two terminating deletion
sequences $\vec{S}= (T, S_1, \ldots, S_m)$, $\vec{S}'= (T,
S'_1, \ldots, S'_{m'})$ 
such that $S_{m} \neq S'_{m'}$; otherwise, we say that $T$ has a
\emph{unique terminating set}.
\pro  No (finite) set of strategy profiles has ambiguous terminating sets.
\epro

\prf
Let $T$ be a set of strategy profiles of least cardinality that
has ambiguous terminating deletion sequences $\vec{S} = (T, S_1,
\ldots, S_m)$ and  $\vec{S}'= (T, S'_1, \ldots, S'_{m'})$, where $S_m
\ne S'_{m'}$.  
Let $T'$ be the set of strategies 
that are not minimax dominated with respect to $T$.
Clearly $T' \ne \emptyset$ and, by definition, $T' \subseteq S_1 \inter
S_1'$.  Since $T'$, $S_1$, and $S_1'$ all have cardinality less than that
of $T$, they must all have unique terminating sets; moreover, the
terminating sets must be the same.  For consider a terminating deletion
sequence starting at $T'$.  We can get a terminating deletion sequence
starting at $S_1$ by just appending this sequence to $S_1$ (or taking
this sequence itself, if $S_1 = T'$).  We can similarly get a terminating
deletion sequence starting at $S_1'$.  Since all these terminating
deletion sequences have the same final element, this must be the unique
terminating set.  But $(S_1, \ldots, S_m)$ and $(S_1', \ldots, S'_{m'})$
are terminating deletion sequences with $S_m \ne S'_{m'}$, a contradiction.
\eprf

\cor The set of strategies that survives iterated deletion of minimax
dominated strategies is independent of the deletion order.
\ecor

\begin{remark}
\label{remark1}
Note that in the definition of $\NSD^k_i(\Gamma)$, we remove all
strategies that 
are dominated by some strategy in $\Sigma_{i}(\Gamma)$, not
just those dominated by some strategy in
$\NSD^{k-1}_{i}(\Gamma)$. Nevertheless,  
the definition would be equivalent even if we had
considered only dominating strategies in
$\NSD^{k-1}_i(\Gamma)$. For suppose not. 
Let $k$ be the smallest integer
such that there exists some strategy $\sigma_i \in
\NSD^{k-1}_{i}(\Gamma)$ that is minimax
dominated by a strategy $\sigma'_i \notin \NSD^{k-1}_i(\Gamma)$, but
there is no strategy in $\NSD^{k-1}_i(\Gamma)$ that dominates $\sigma_i$.
That is, $\sigma'_i$ was deleted in some previous iteration.
Then there exists a sequence of strategies $\sigma^0_i, \ldots,
\sigma^q_i$ and indices $k_0 < k_1 < \ldots < k_q = k-1$ such that
$\sigma^0_i = \sigma'_i$, $\sigma^j_i \in NSD_i^{k_j}(\Gamma)$, 
and for all $0\leq j < q$, $\sigma_i^{j}$ 
is minimax dominated by $\sigma_i^{j+1}$ with respect to
$\NSD_i^{k_j-1}(\Gamma)$.  
Since $\NSD^{k-2}(\Gamma) \subseteq \NSD^{j}(\Gamma)$ for $j \le k-2$,
an easy induction on $j$ shows that $\sigma_i^q$ minimax dominates
$\sigma^{q-j}$ with respect to $\NSD^{k-2}$ for all $0 < j \le q$.
In particular, $\sigma^q$ minimax dominates $\sigma_i^0 = \sigma'$ with
respect to $\NSD^{k-2}$.
\qed
\end{remark}

The following example shows that iteration has bite: there exist a
2-player game where each player has $k$ actions and $k-1$ rounds
of iterations are needed.
\begin{example}
\label{ex1} Consider a two-player game, where both players
  announce a value between 1 and $k$. Both players receive in utility
  the smallest of the values announced; additionally, the player who
  announces the larger value get a reward of $p= 0.5$.\footnote{This
    game can be viewed a a reverse variant of the 
Traveler's dilemma \cite{Basu94}, where the player who announces the
  smaller value gets the 
  reward.} That is, 
 $u(x,y) = (y+p,y)$ if $x > y$, $(x,x+p)$ if $y > x$, and $(x,x)$ if $x=y$.
In the first step of the deletion process, 1 is deleted for both
players; playing $1$ can yield a max utility of $1$, whereas the minimum
utility of any other action is $1.5$. Once 1 is deleted, 2 is deleted
for both players: 2 can yield a max utility of 2, and the min utility
of any other action (once 1 is deleted) is 2.5.
Continuing this process, we see that only $(k,k)$ survives.
\qed
\end{example}

\subsection{Characterizing Iterated Minimax Domination}
We now show that strategies surviving iterated removal of minimax
dominated strategies characterize the set of strategies consistent
with common counterfactual belief of rationality in (strongly)
appropriate counterfactual structures. 
As a first step, we define a ``minimax'' analogue of
rationalizability. 
\dfn \label{rat2} A strategy profile 
$\vec{\sigma}$ in game~$\Gamma$  
is \emph{minimax rationalizable} if, for each player $i$, there is a set $\Z_i   
\subseteq \Sigma_i(\Gamma)$ such that 
\begin{itemize}
\item $\sigma_i \in \Z_i$;
\item for every strategy $\sigma'_i \in \Z_i$ and strategy $\sigma''_i
  \in \Sigma_{i}(\Gamma)$,
$$\max_{\tau_{-i} \in \Z_{-i}} u_i(\sigma'_i,\tau_{-i}) \geq
\min_{\tau_{-i} \in \Z_{-i}} u_i(\sigma''_i,\tau_{-i}).$$ 
\end{itemize}
\edfn

\thm\label{thm:NSD} 
The following are equivalent: 
\begin{enumerate}
\item[(a)] $\vec{\sigma} \in \NSD^{\infty}(\Gamma)$;
\item[(b)] $\vec{\sigma}$ is minimax rationalizable in
  $\Gamma$;
\item[(c)] there exists a finite counterfactual structure $M$
that is strongly 
appropriate for $\Gamma$ and a state
$\omega$ such that 
$(M,\omega) \sat \play(\vec{\sigma}) \land_{i =1}^n \SRAT_i^k$ for
all $k \geq 0$; 

\item[(d)] for all players $i$, there exists a finite counterfactual
structure $M$ 
that is 
appropriate for $\Gamma$ and a state $\omega$
such that 
$(M,\omega) \sat \play_i(\sigma_i) \land \SRAT_i^k$ for all $k \geq 0$.
\end{enumerate}
\ethm
%
%
%
%
%
%
%
%
\prf
We prove that (a) implies (b) implies (c) implies (d) implies (a).
We first introduce some helpful notation.
Recall that $\arg\max_x f(x) = \{y: \mbox{ for all }
 z,  f(z) \le f(y)\}$; $\arg\min_x f(x)$ is defined similarly.
 For us, $x$ ranges over pure strategies or pure strategy
profiles, and we will typically be interested in considering some
element of the set, rather than the whole set.  Which element we take
does not matter.  We thus assume that there is some order on the set of
pure strategies and strategy profiles, and take the $\arg\max^*_x f(x)$
to be the maximum element of $\arg\max_x f(x)$ with respect to this
order; $\arg\min^*_x f(x)$ is defined similarly.

\paragraph{(a) $\Rightarrow$ (b):} 
Let $K$ be an integer such that $\NSD^K(\Gamma) = \NSD^{K+1}(\Gamma)$;
such a $K$ must exist since the game is finite. It also easily follows
that for each player $j$, $\NSD^{K}_j(\Gamma)$ is non-empty: in 
iteration $k+1$, no $\NSD^k_j$-maximin strategy, that is, no strategy in
$\arg\max_{\sigma'_j \in 
  \NSD^{k}_j(\Gamma)} \min_{\tau_{-j} \in \NSD^k_j(\Gamma)}
  u_j(\sigma'_j, \tau_{-j})$, is deleted, since no maximin strategy is
 minimax  dominated by a strategy in    $\NSD^k_j(\Gamma)$ (recall that by
  Remark \ref{remark1}, it suffices to consider domination by strategies
  in $\NSD^{k}_j(\Gamma)$).
 Let $\Z'_j = \NSD^K_j(\Gamma)$. It immediately follows that the sets $\Z'_1,
\ldots, \Z'_n$ satisfy the conditions of Definition \ref{rat2}.

\paragraph{(b) $\Rightarrow$ (c):} 
Suppose that $\vec{\sigma}$ is minimax rationalizable.  Let 
$\Z_1, \ldots, \Z_n$ be the sets guaranteed to exist by
Definition \ref{rat2}.
Let $\W^i = \{(\vec{\sigma}, i)  \; | \; \vec{\sigma} \in \Z_{-i} \times
\Sigma_{i}\}$, and let $\W^0 = 
\{(\vec{\sigma}, 0)  \; | \; \vec{\sigma} \in \Z_1
\times \ldots \times \Z_n\}$.
Think of $\W^0$ as states where
everyone is (higher-level) rational, and 
of $\W^i$ as ``counterfactual'' states where player $i$ has
changed strategies.
In states in $\W^0$, each player $j$ assigns probability 1 to the
other players choosing actions that \emph{maximize} $j$'s utility (given
his action). On the other hand, in states in $\W^i$, where $i \neq 0$,
player $i$ assigns 
probability 1 to the other players choosing actions that \emph{minimize}
$i$'s utility, whereas all other player $j \neq i$ still assign
probability 1 to other players choosing actions that maximize
$j$'s utility.

Define a structure 
$M = (\Omega,f,\strat, \PR_1, \ldots, \PR_n)$, where
\begin{itemize}
\item $\Omega = \cup_{i \in \{0, 1, \ldots, n\}} \W^i$;
\item $\strat(\vec{\sigma}',i) = \vec{\sigma}'$;
\item $\PR_j(\vec{\sigma}',i)(\vec{\sigma}'',i') =
\left\{\begin{array}{lll}
1 &\mbox{if $i =j = i'$, $\sigma'_i =
\sigma''_i$, and $\sigma''_{-i} = \arg\min^*_{\tau_{-i} \in
  \Z_{-i}}u_j(\sigma'_i,\tau_{-i})$,} \\
1 &\mbox{if $i\neq j$, $i' = 0$, and $\sigma'_j =
\sigma''_j$, and $\sigma''_{-j} = \arg\max^*_{\tau_{-j} \in
  \Z_{-j}}u_j(\sigma'_j,\tau_{-j})$,} \\
0 &\mbox{otherwise;}
\end{array}
\right.
$
\item $f((\vec{\sigma}',i),j,\sigma''_j) = \left\{
\begin{array}{lll}
(\vec{\sigma}',i) &\mbox{if $\sigma'_j =
  \sigma''_j$,}\\
((\sigma''_j,\tau'_{-j}), j) &\mbox{otherwise, where 
$\tau'_{-j} = \arg\min^*_{\tau_{-j} \in \Z_{-j}}
u_j(\sigma'_j,\tau_{-j})$.}\\
\end{array}\right.
$
\end{itemize}
It follows by inspection that $M$ is 
strongly 
appropriate
for $\Gamma$.
We now prove by induction on $k$ that, for all $k \geq 1$
all $i \in \{0,1, \ldots, n\}$, and all states $\omega \in \W^i$,
$(M,\omega) \sat \land_{j \neq i}  \SRAT^k_j$.
For the base case $(k=1)$, since $\SRAT^1_j$ is logically equivalent to
$\RAT_j$, we must show that if $\omega \in \W^i$,
then $(M,\omega) \sat \land_{j \neq i}  \RAT_j$.
Suppose that $\omega = (\vec{\sigma}',i) \in \W^i$.  If $i \ne j$,  
then at $\omega$, player $j$ places probability 1 on the true state
being $\omega' = (\vec{\sigma}'',0)$, where $\sigma''_j = \sigma'_j$ 
and $\sigma''_{-j} = \arg\max^*_{\tau_{-j} \in \Z_{-j}}
u_j(\sigma'_j,\tau_{-j})$.  Player $j$ must be rational, since 
if there exists some strategy $\tau'_j$ such that
$u_j(\vec{\sigma}'') < \sum_{\omega' \in \Omega}
\cPR_{j,\tau'_j}(\omega)(\omega') u_j( \tau'_j, 
\strat_{-j}(\omega'))$, then the definition of $\PR_j$ guarantees that
$u_j(\vec{\sigma}'') < u_j(\tau_j',\vec{\tau}''_{-j})$, where 
$\tau''_j = \arg\min^*_{\tau_{-j} \in \Z_{-j}}
u_j(\sigma'_j,\tau_{-j})$.  If this inequality held, then $\tau'_j$
would minimax dominate $\sigma'_j$, contradicting the assumption that
$\sigma'_j \in \Z_j$.    
%
For the induction step, suppose that the result holds for $k$; we show 
that it holds for $k+1$. 
Suppose that $\omega \in \W^i$ and $j \ne i$.  By construction,
the support of $\PR_{j}(\omega)$ is a subset of $\W^0$; by the
induction hypothesis, it follows that
$(\M,\omega) \sat \B_j (\land_{j' =1}^n \SRAT^{k}_{j'})$.
Moreover, by construction, it follows that for all players $j$ and 
all strategies $\sigma'_j \neq \strat_i(\omega)$, the support of
$\cPR_{j,\sigma'_j}(\omega)$ is a subset of $\W^j$. By the
induction hypothesis, it follows that for all $j
\neq i$, $(\M,\omega) \sat \K_j (\land_{j' \neq j} \SRAT^{k}_{j'})$.
Finally, it follows from the induction hypothesis that for all $j \neq i$, 
$(\M,\omega) \sat \SRAT^{k}_{j}$. Since $\SRAT^k_j$ implies
$\RAT_j$, it follows that for all $j \neq i$, 
$(\M,\omega) \sat \RAT_j \land \K_j (\land_{j' \neq j}
\SRAT^{k}_{j'})$, which proves the induction step.

\paragraph{(c) $\Rightarrow$ (d):} The implication is trivial.

\paragraph{(d) $\Rightarrow$ (a):}
We prove an even stronger statement: For all $k\geq 0$, if there
exists a finite counterfactual structure $\M^k$
that is 
appropriate for $\Gamma$ and a state $\omega$
such that 
$(\M^k,\omega) \sat \play_i(\sigma_i) \land \SRAT^k_i$, then $\sigma_i\in
\NSD_i
^k(\Gamma)$.\footnote{The
  converse also holds; we omit the details.}
We proceed by induction on $k$.
The result clearly holds if $k=0$.  
Suppose that the result holds for $k-1$ for $k \ge 1$; we show that it
holds for $k$. 
Let $\M^k=(\Omega,f,\strat, \P_1, \ldots, \P_n)$ be a finite counterfactual
structure that is 
appropriate for $\Gamma$ 
and a state $\omega$ such that 
$(\M^k,\omega') \sat
\play_i(\sigma_i) \land \SRAT_i^{k}$.
Replacing $\SRAT_i^k$ by its definition, we get that 
$$(\M^k,\omega') \sat
\play_i(\sigma_i) \land \RAT_i\land \K_i (\land_{j \neq i} \SRAT_j^{k-1}).$$
By definition of $\K_i$, it follows that 
for all strategies $\sigma'_i$ for player $i$ and all $\omega''$ such
that $\cPR_{i,\sigma'_i}(\omega')(\omega'') > 0$, 
$$(\M^k,\omega'') \sat
\land_{j \neq i} \SRAT_j^{k-1},$$ so by the induction hypothesis, 
it follows that for all $\omega''$ such that
$\cPR_{i,\sigma'_i}(\omega')(\omega'') > 0$, we have 
$\strat_{-i}(\omega'') \in \NSD_{-i}^{k-1}(\Gamma)$. 
Since $(\M^k,\omega') \sat
\play_i(\sigma_i) \land \RAT_i$, 
it follows that $\sigma_i$ cannot be minimax dominated
with respect to $\NSD_{-i}^{k-1}(\Gamma)$. Since, for
all $j'>1$, $\NSD_{-i}^{j'}(\Gamma) \subseteq
\NSD_{-i}^{j'-1}(\Gamma)$, it follows that, for all $k' < k$, $\sigma_i$
is not minimax  dominated with respect to $\NSD_{-i}^{k'}(\Gamma)$.
Thus,
$\sigma_i \in \NSD_i^{k}(\Gamma)$.
\eprf

\section{Characterizing Analogues of Nash Equilibrium}
\label{NE.sec}
In this section, we consider analogues of Nash equilibrium
in our setting. This allows us to 
relate our approach to the
work of Tennenholtz \citeyear{tennenholtz} and Kalai et al.~\citeyear{KKLS}.
In the standard setting, if a strategy profile
$\vec{\sigma}$ is a
Nash equilibrium, then there exists a state where $\vec{\sigma}$ is
played, common belief of rationality holds, and additionally, the
strategy profile is (commonly) known to the players.
To study analogues of Nash equilibrium, we thus investigate the effect of
adding assumptions about knowledge of the players'
strategies.
We consider several ways of formalizing this.
The weakest approach is to simply require that the actual
strategies used by the players is known.
\begin{itemize}
\item $(M,\omega) \sat \MKS$ iff, for all players $i$, 
$$\PR_i(\omega)(\intension{\strat_{-i}(\omega)}) 
= 1.$$
\end{itemize}
$\MKS$ does not require that player $i$ knows how players $-i$ will respond
to $i$ switching strategies. A stronger condition would be to require
not only that every player $i$ knows the strategies of the other players,
but also how they respond to $i$ switching strategies.
\begin{itemize}
\item $(M,\omega) \sat \MKR$ iff, for all players $i$ and
strategies $\sigma'_i$ for $i$,  
$$\cPR_{i,\sigma'_i}(\omega)(\intension{\strat_{-i}(f(\omega,i,\sigma'_i))}) 
= 1.$$
\end{itemize}
Clearly, $\MKR$ implies $\MKS$ (by simply considering $\sigma'_i =
\strat_i(\omega)$).
An even stronger condition is to require that the players
know the true state of the world.
\begin{itemize}
\item $(M,\omega) \sat \MKW$ iff, for all players $i$,
$$\PR_{i}(\omega)(\omega) = 1.$$
\end{itemize}
Note that if all players know the true state of the
world, then they also counterfactually know the true state of the
world: for every player $i$ and every strategy $\sigma'_i$ for player $i$,  
$$\cPR_{i,\sigma'_i}(\omega)(f(\omega,i,\sigma'_i)) = 1.$$
It follows that $\MKW$ implies $\MKR$ and thus also $\MKS$.
Additionally, note that $\MKW$ implies $\EB (\MKW)$, so $\MKW$ also
implies $\CB (\MKW)$. 

We now characterize $\CKR$ in structures satisfying
the conditions above.
We say that a strategy profile $\vec{\sigma}$ is \emph{individually rational}
(IR) if for every player $i$ in the game $\Gamma$, 
$$u_i(\vec{\sigma}) \geq \max_{\sigma'_i \in \Sigma_i(\Gamma)} \min_{\tau_{-i}\in
\Sigma_{-i}(\Gamma)} u_i(\sigma',\tau_{-i}).$$
Although every IR strategy profile is contained in
$\NSD^1(\Gamma)$, it is not necessarily contained in
$\NSD^2(\Gamma)$. That is, IR strategies may not survive 
two rounds of deletion of minimax dominated strategies.  To see this,
consider the game $\Gamma$ in Example \ref{ex1}. Both players' maximin
payoff is $1.5$, so every strategy profile in $\NSD^1(\Gamma) =
\{ (x,y) 
\; \; | \; \; 2\leq x,y\leq k \}$ is IR, but $\NSD^2(\Gamma)$ does not
contain $(2,2)$.

As the following simple example shows, not every
strategy profile that survives deletion iterated deletion of minimax
dominated strategies is IR.
\begin{example} 
\label{ex2}
Consider the game with payoffs given in the table below.
\begin{table}[h]
\begin{center}
\begin{tabular}{c |  c c}
& $c$ & $d$\\
\hline
$a$ &$(100,0)$ &$(100,0)$ \\
$b$  &$(150,0)$  &$(50,0)$  \\
\end{tabular}
\end{center}
\end{table}

\noindent All strategy profiles survive iterated deletion of minimax dominated
strategies, but $(b,d)$ is not individually rational since playing
$a$ always guarantees the row player utility 100. 
\qed
\end{example}

Let $IR(\Gamma)$ denote the set of IR strategy profiles in $\Gamma$,
and let $IR(\Z_1 \times \ldots \times \Z_n, \Gamma)= IR(\Gamma')$ where $\Gamma'$
is the subgame of $\Gamma$ obtained by restricting player $i$'s strategy
set to $\Z_i$. That is, $IR(\Z_1 \times \ldots \times
\Z_n, \Gamma)$ is the set of strategies $\vec{\sigma} \in \Z_1 \times
\ldots \times \Z_n$ such that for every player $i$, 
$$u_i(\vec{\sigma}) \geq \max_{\sigma'_i \in \Z_i} \min_{\tau_{-i}\in
\Z_{-i}} u_i(\sigma',\tau_{-i}).$$
A stronger way of capturing individual rationality of subgames 
is to require that the condition above hold even if the max is 
taken over every $\sigma'_i \in \Sigma(\Gamma)$ (as opposed to only
$\sigma'_i \in \Z_i$). More precisely, let $IR'(\Z_1 \times \ldots \times
\Z_n, \Gamma)$ be the set of strategies $\vec{\sigma} \in \Z_1 \times
\ldots \times \Z_n$ such that, for all players $i$, 
$$u_i(\vec{\sigma}) \geq \max_{\sigma'_i \in \Sigma_i(\Gamma)}
\min_{\tau_{-i}\in 
\Z_{-i}} u_i(\sigma',\tau_{-i}).$$

Our characterization of $\CKR$ in the presence of (common)
knowledge of strategies follows.

\thm\label{thm:NSD1} 
The following are equivalent: 
\begin{enumerate}
\item[(a)] $\vec{\sigma} \in IR(\NSD^{\infty}(\Gamma), \Gamma)$;
\item[(b)] $\vec{\sigma} \in IR'(\NSD^{\infty}(\Gamma), \Gamma)$;
\item[(c)] $\vec{\sigma}$ is minimax rationalizable 
and $\vec{\sigma} \in IR'(\Z_1 \times \ldots \times
\Z_n,\Gamma)$, where   $\Z_1, \ldots, \Z_n$ are 
the sets of strategies   guaranteed to exists by the definition of minimax
rationalizability; 
\item[(d)] there exists a finite counterfactual structure $M$
that is strongly 
appropriate for $\Gamma$ and a state $\omega$ such that
$(M,\omega) \sat \MKW \land \play(\vec{\sigma}) \land_{i=1}^n
\SRAT_i^k$ for every $k \geq 0$; 
\item[(e)]
there exists a finite counterfactual structure $M$
that is 
appropriate for $\Gamma$ and a state $\omega$
such that
$(M,\omega) \sat \MKS \land \play(\vec{\sigma}) \land_{i =1}^n
\SRAT_i^k$ for every $k \geq 0$.
\end{enumerate}
\ethm
\prf
Again, we prove that (a) implies (b) implies (c) implies (d) implies
(e) implies (a).
\paragraph{(a) $\Rightarrow$ (b):}
We show that if $\vec{\sigma} \in IR(\NSD^k(\Gamma), \Gamma)$ then 
$\vec{\sigma} \in IR'(\NSD^k(\Gamma), \Gamma)$. The implication then
follows from the fact that since the game is finite there exists some
$K$ such that $\NSD^K(\Gamma) = \NSD^{\infty}(\Gamma)$.

Assume by way of contradiction that $\vec{\sigma} \in IR(\NSD^k(\Gamma),
\Gamma)$ but  
$\vec{\sigma} \notin IR'(\NSD^k(\Gamma), \Gamma)$; that is, 
there exists a player $i$ and a strategy
$\sigma'_i \notin \NSD_i^{k}(\Gamma)$ such that 
$$\min_{\tau_{-i}\in \NSD_{-i}^k(\Gamma)} u_i(\sigma'_i,\tau_{-i}) >
u_i(\vec{\sigma}). $$ 
By the argument in Remark \ref{remark1}, there exists a strategy
$\sigma''_i \in \NSD^k_i(\Gamma)$ such that 
$u_i(\sigma''_i, \tau''_{-i}) >  u_i(\sigma'_i, \tau'_{-i})$ for
all $\tau''_{-i}, \tau'_{-i} \in NSD_{-i}^k(\Gamma)$.
It follows that 
$$\min_{\tau_{-i}\in \NSD_{-i}^k(\Gamma)} u_i(\sigma''_i,\tau_{-i}) >
u_i(\vec{\sigma}). $$
Thus, $\vec{\sigma} \notin IR(\NSD^k(\Gamma), \Gamma)$.

\paragraph{(b) $\Rightarrow$ (c):} 
The implication follows in exactly the same way as in 
the proof that (a) implies (b) in
Theorem \ref{thm:NSD}.

\paragraph{(c) $\Rightarrow$ (d):} 
Suppose that $\vec{\sigma}$ is minimax rationalizable.  Let 
$\Z_1, \ldots, \Z_n$ be the sets guaranteed to exist by
Definition \ref{rat2}, and suppose that $\vec{\sigma} \in IR'(\Z_1 \times
\Z_n, \Gamma)$. Define the sets $\W^i$ as in the proof
of Theorem \ref{thm:NSD}.
Define the structure $M$ just as in the proof 
of Theorem \ref{thm:NSD}, except that 
for all players $i$, let
$\PR_i((\vec{\sigma},0))((\vec{\sigma}',i')) = 
1$ in case $\vec{\sigma}' = \vec{\sigma}$ and $i' = 0$.
Clearly $(M,(\vec{\sigma},0)) \sat \MKW$.
It follows using the same arguments as in the proof of Theorem
\ref{thm:NSD} that $M$ is 
strongly 
appropriate and
that 
$(M,(\vec{\sigma},0) \sat \play(\vec{\sigma}) \land_{i=1}^n
\SRAT_i^k$ for 
every $k \geq 0$; we just need to rely on the (strong) IR property of
$\vec{\sigma}$ to prove the base case of the induction.

\paragraph{(d) $\Rightarrow$ (e):} The implication is trivial.

\paragraph{(e) $\Rightarrow$ (a):}
Recall that since the game is finite, there exists a constant $K$
such that $NSD^{K-1}(\Gamma) = NSD^{K}(\Gamma) = \NSD^{\infty}(\Gamma)$.
We show that if there
exists a finite counterfactual structure $M$
that is 
appropriate for $\Gamma$ and a state $\omega$ such that
$(M,\omega) \sat \MKS \land \play(\vec{\sigma}) \land_{i=1}^n
\SRAT_i^K$, then $\vec{\sigma} \in
IR(\NSD^K(\Gamma), \Gamma)$.

Consider some state $\omega$ such that 
$(M,\omega) \sat \MKS \land \play(\vec{\sigma}) \land_{i =1}^n
\SRAT_i^K$.
By Theorem \ref{thm:NSD}, it follows that $\vec{\sigma} \in \NSD^K(\Gamma)$.
For each player $i$, it additionally follows that 
$(M,\omega) \sat \play(\vec{\sigma}) \land \EB (\play(\vec{\sigma}))
\land \RAT_i \land \K_i (\land_{j \neq i} \SRAT_j^{K-1})$.
By Theorem \ref{thm:NSD}, it follows that for every strategy
$\sigma'_i$ for $i$, the support of the projection of
$\cPR_{i,\sigma'_i}(\omega)$ onto strategies for players $-i$ is a
subset of $\NSD_{-i}^{K-1}(\Gamma) = \NSD_{-i}^{K}(\Gamma)$.
Thus, we have that for every $\sigma'_i$, there exists $\tau_{-i} \in
\NSD^{K}_{-i}(\Gamma)$ such that $u_i(\vec{\sigma}) \geq
u_i(\sigma'_i, \tau_{-i})$, which means that $\vec{\sigma}$ is IR in
the subgame induced by restricting the strategy set to $\NSD^K(\Gamma)$.
\eprf

It is worth comparing Theorem \ref{thm:NSD1} to the results of Tennenholtz
\citeyear{tennenholtz} and Kalai et al.~\citeyear{KKLS} on program
equilibria/equilibria with conditional 
commitments.
Recall that these papers focus on 2-player games.  In Tennenholtz's
model, each player $i$ deterministically picks a
program $\Pi_i$; player $i$'s action is $\Pi_i(\Pi_{-i})$. In the
two-player case, a
program equilibrium is a pair of programs $(\Pi_1,\Pi_2)$ such that
no player can improve its utility by unilaterally changing its
program. In this model any IR strategy profile $(a_1,a_2)$ can be
sustained in a program equilibrium: each player uses the program $\Pi$, where
$\Pi(\Pi')$ outputs $a_i$ if $\Pi' = \Pi$, and otherwise ``punishes''
the other player using his minmax strategy.
(Tennenholtz extends this result to show that any 
mixed IR strategy profile can be sustained in a program equilibrium, by
considering randomizing programs; Kalai et al.~show that all correlated
IR strategy profiles can be sustained, by 
allowing the players to pick a distribution over programs.)
In contrast, in our model, a smaller set of strategy profiles can be
sustained. This difference can be explained as follows.
In the program equilibrium model a player may ``punish'' the other
player using an arbitrary action (e.g., using minimax punishment)
although
this may be detrimental for him. Common counterfactual belief of
rationality disallows such punishments. More precisely, it 
allows a player $i$ to punish other players only by using a strategy that is
rational for player $i$. On the other hand, 
as we now show, 
if we require only common belief (as opposed to
counterfactual belief) in rationality, then
any IR strategy can be sustained in an equilibrium in our model.

\thm\label{thm:IR} 
The following are equivalent: 
\begin{enumerate}
\item[(a)] $\vec{\sigma} \in IR(\Gamma)$;
\item[(b)] there exists a finite counterfactual structure $M$
that is strongly 
appropriate for $\Gamma$ and a state $\omega$ such that
$(M,\omega) \sat \MKW \land \play(\vec{\sigma}) \land \CB (\RAT)$;
\item[(c)] there exists a finite counterfactual structure $M$
that is 
appropriate for $\Gamma$ and a state $\omega$
such that
$(M,\omega) \sat \MKS \land \play(\vec{\sigma}) \land \CB (\RAT)$.
\end{enumerate}
\ethm
\prf
Again, we prove that (a) implies (b) implies (c) implies (a).

\paragraph{(a) $\Rightarrow$ (b):} 
Define a structure 
$M = (\Omega,f,\strat, \PR_1, \ldots, \PR_n)$, where
\begin{itemize}
\item $\Omega = \Sigma(\Gamma)$; 
\item $\strat(\vec{\sigma}') = \vec{\sigma}'$;
\item $\PR_j(\vec{\sigma}')(\vec{\sigma}') = 1$.
\item $f(\vec{\sigma}',i,\sigma''_j) = \left\{
\begin{array}{lll}
\vec{\sigma}' &\mbox{if $\sigma'_j =
  \sigma''_j$,}\\
(\sigma''_j,\tau'_{-j}) &\mbox{otherwise, where 
$\tau'_{-j} = \arg\min^*_{\tau_{-j} \in \Sigma_{-j}(\Gamma)}
u_j(\sigma'_j,\tau_{-j})$.}
\end{array}\right.$
\end{itemize}
It follows that $M$ is strongly 
appropriate for
$\Gamma$ and that $(M,\vec{\sigma}) \sat \MKW$. 
Moreover, $(M,\vec{\sigma}) \sat RAT$ since 
$\vec{\sigma}$ is individually rational; furthermore, since each player
considers only the 
state $\vec{\sigma}$ possible at $\vec{\sigma}$, it follows that 
$(M,\vec{\sigma}) \sat \CB (RAT)$. 

\paragraph{(b) $\Rightarrow$ (c):} The implication is trivial. 

\paragraph{(c) $\Rightarrow$ (a):} 
Suppose that $M = (\Omega,f, \strat, \PR_1, \ldots, \PR_n)$ is
a finite counterfactual structure appropriate for $\Gamma$, and 
$(M,\omega) \sat \MKW \land \play(\vec{\sigma})\land \CB (\RAT)$.  
It follows that for each player $i$, 
$(M,\omega) \sat \play(\vec{\sigma}) \land \EB (\play(\vec{\sigma}))
\land \RAT_i$.
Thus, we have that for all strategies $\sigma'_i$, there exists $\tau_{-i} \in
\Sigma_{-i}(\Gamma)$ such that $u_i(\vec{\sigma}) \geq
u_i(\sigma'_i, \tau_{-i})$, which means that $\vec{\sigma}$ is IR.
\eprf

\section{Discussion}
We have introduced a game-theoretic framework for analyzing scenarios
where a player may believe that if he were to switch strategies, this
intention to switch may be detected by the other players, resulting in
them also switching strategies. Our formal model allows players'
counterfactual beliefs (i.e., their beliefs about the state of the
world in the event that they switch strategies) to be arbitrary---they
may be completely different from the players' actual beliefs.

We may also consider a more restricted model where we require 
that
a player $i$'s 
counterfactual beliefs 
regarding other players' strategies and beliefs 
is
$\epsilon$-close to player $i$'s actual beliefs in
total variation distance\footnote{Recall that two probability
  distribution are $\epsilon$-close 
  in total variation distance if the probabilities that they assign to any
  event $E$ differ by at most $\epsilon$.}---that is, for every state $\omega \in \Omega$, player $i$, and
strategy $\sigma'_i$ for player 
$i$, 
%
%
the projection of $\cPR_{i,\sigma'_i}(\omega)$ onto strategies and beliefs
of players $-i$ is $\epsilon$-close to the projection of
$\PR_i(\omega)$ onto strategies and beliefs of players $-i$.

We refer to counterfactual 
structures satisfying this property as $\epsilon$-counterfactual
stuctures. Roughly speaking, $\epsilon$-counterfactual structures
restrict to scenarios where players are not ``too'' transparent to one
another; this captures the situation when a player assigns only
some ``small'' probability to its switch in action being noticed by the
other players.
$0$-counterfactual structures behave just as counterfactual structures
that respect 
unilateral deviations:
common counterfactual belief of rationality in $0$-counterfactual
structures characterizes 
rationalizable strategies (see Remark \ref{unilateral.rem}). 
The general counterfactual structures investigated in this paper are
$1$-counterfactual structures
(that is, we do not impose any conditions on players' counterfactual
beliefs). 
We remark that although our characterization 
results rely on the fact that we consider $1$-counterfactual
structures, the motivating example in the introduction (the
translucent prisoner's dilemma game) shows that even considering 
$\epsilon$-counterfactual structures with a small $\epsilon$ can result
in there being strategies
consistent with common counterfactual belief of rationality 
that are not rationalizable.
We leave an exploration of 
common counterfactual belief 
of rationality in
$\epsilon$-counterfactual structures for future work.

\bibliographystyle{chicagor}
\bibliography{z,joe}

\appendix

\end{document}